# Atmospheric constituents and surface-level UVB: implications for a paleoaltimetry proxy and attempts to reconstruct UV exposure during volcanic episodes

Brian C. Thomas[1*], Byron D. Goracke[1], and Sean M. Dalton[1]

[1]Washburn University, Department of Physics and Astronomy, Topeka, KS;

[*]Corresponding author: 1700 SW College Ave., Topeka, KS 66604; 785-670-2144;

brian.thomas@washburn.edu

Abstract:

Chemical and morphological features of spores and pollens have been linked to changes in solar ultraviolet radiation (specifically UVB, 280-315 nm) at Earth's surface. Variation in UVB exposure as inferred from these features has been suggested as a proxy for paleoaltitude; such proxies are important in understanding the uplift history of high altitude plateaus, which in turn is important for testing models of the tectonic processes responsible for such uplift. While UVB irradiance does increase with altitude above sea level, a number of other factors affect the irradiance at any given place and time. In this modeling study we use the TUV atmospheric radiative transfer model to investigate dependence of surface-level UVB irradiance and relative biological impact on a number of constituents in Earth's atmosphere that are variable over long and short time periods. We consider changes in $O_3$ column density, and $SO_2$ and sulfate aerosols due to periods of volcanic activity, including that associated with the formation of the Siberian Traps. We find that UVB irradiance may be highly variable under volcanic conditions and variations in several of these atmospheric constituents can easily mimic or overwhelm changes in UVB irradiance due to changes in altitude. On the other hand, we find that relative change with altitude is not very sensitive to different sets of atmospheric conditions. Any paleoaltitude proxy based on UVB exposure requires confidence that the samples under comparison were located at roughly the same latitude, under very similar $O_3$ and $SO_2$ columns, with similar atmospheric aerosol conditions. In general, accurate estimates of the surface-level UVB exposure at any time and location require detailed radiative transfer modeling taking into account a number of atmospheric factors; this result is important for paleoaltitude proxies as well as attempts to reconstruct the UV environment through geologic time and to tie extinctions, such as the end-Permian mass extinction, to UVB irradiance changes.

## 1. Introduction

There has been interest in the geology and paleontology literature regarding variations of solar ultraviolet-B (UVB) radiation at Earth's surface through geologic time (Harfoot et al., 2007; Willis et al., 2009; Fraser et al., 2014). UVB is biologically active so variations due to changes in environmental conditions may be important in understanding the history of life on Earth. Indeed, depletion of stratospheric ozone ($O_3$) has been suggested as a factor in mass extinctions such as the end-Permian (Beerling et al., 2007; Meyer and Kump, 2008).

Several studies have found morphological and biochemical features in spores and pollen that are affected by UVB, which may serve as a proxy in the fossil record for changes in UVB exposure (Yeloff et al., 2008; Willis et al. 2011; Fraser et al., 2014). These proxies may allow estimates of a number of environmental changes in Earth's history. Lomax et al. (2012) propose using a chemical proxy for ultraviolet-B (UVB) exposure as a proxy for paleoaltitude.

Paleoaltitude proxies are important in understanding the uplift history of high altitude plateaus, which in turn is important for testing models of the tectonic processes responsible for such uplift. According to Lomax et al. (2012, pg. 22), "competing models... could be rigorously assessed if uplift histories of plateaus were tightly constrained" and "quantitative paleoproxies to determine the paleoaltimetry of high altitude plateaus such as the Tibetan Plateau [are] of interest to a wide spectrum of scientists working within the geoscience community."

Given the importance of understanding changes in UVB we have carried out a study to investigate absolute and relative changes in irradiance caused by changes in atmospheric $O_3$, $SO_2$, and sulfate aerosols. We examine UVB changes in the context of the Lomax et al. (2012) paleoaltimetry proxy, but our results will be useful for anyone seeking to understand or predict variations in surface-level UVB.

Lomax et al. (2012) discuss evidence that chemical properties of the walls of spores and pollen are affected by UVB exposure, with UVB defined as 280-315 nm (see also Fraser et al., 2014 and Yeloff et al., 2008). There is ambiguity in using irradiance integrated across the 280-315 nm waveband as a direct measure of biological impact. Typically, effects of UVB are quantified using a "biological weighting function" (BWF; also known as an "action spectrum"), taking into account the wavelength-dependent response, giving a weighted irradiance or relative effectiveness. A common weighting function is that for erythema (sunburn) in humans; many others exist, e.g. direct DNA damage and reduced photosynthetic productivity. Wavelength-dependence may be important since atmospheric constituents have absorption bands in different wavelength regions, complicating the connection between changing constituents and biological impact.

A specific weighting function for chemical changes in spores and pollen does not appear to exist. Work in this area has investigated changes observed under varying levels of integrated UVB irradiance, without determining a wavelength-dependent response (Rozema 2009; Willis 2011). On the other hand, Rozema et al. (2001) used the Generalized Plant Action Spectrum of Caldwell (1971) as a weighting function to compute UVB dose for their experimental work. It is plausible that greater potential damage would correspond to greater change in UVB screening properties. Here we consider changes in UVB irradiance and changes in relative biological impact, using the erythema action spectrum of McKinlay and Diffey (1987) and the Generalized Plant Action Spectrum of Caldwell (1971).

UVB increases with altitude, but many other factors can affect the amount of UVB received at a given altitude. UVB at Earth's surface has varied over geologic history (Willis et al., 2009), due to factors including changes in Earth's orbit, varying solar activity and luminosity, and changes in atmospheric constituents.

Our purpose is to examine the impact on UVB at Earth's surface by variations in several atmospheric constituents. We consider variations in ozone column density and volcanic constituents. We focus on changes that could introduce uncertainties into a chemical-UVB proxy used as a paleoaltimetry proxy.

To our knowledge, this is the first such study to utilize full radiative transfer modeling coupled with quantitative biological weighting functions, and to examine a wide range of causes of UVB variation, both independently and simultaneously. Our results should serve as a resource for those interested in variation of UVB exposure for a variety of applications.

## 2. Variations of atmospheric constituents

A variety of atmospheric constituents are important for transmission of UVB through Earth's atmosphere. Here we discuss some of the important factors; for more details see the review articles by Whitehead et al. (2000), DeLuisi (1997), and Madronich (1993).

Ozone is an important absorber of UVB, with a strong absorption band between 200 and 360 nm. Ozone in Earth's atmosphere is concentrated in the stratosphere and produced primarily in the tropics by photolysis of $O_2$ by solar radiation with wavelength less than 240 nm. While production of $O_3$ is highest in the tropics, upwelling and Brewer-Dobson circulation leads to efficient transport toward the poles, leading to higher total column $O_3$ at higher latitudes. Destruction of $O_3$ occurs by direct photolysis (by photons with wavelength less than 410 nm) as well as by reactions with various chemical constituents. A general background level of $O_3$ is achieved, varying in space and time depending on changes in solar radiation, atmospheric chemistry, and transport.

Major changes in $O_3$ can be caused by transient events such as volcanic eruptions (Rosenfield et al., 1997), major solar activity (Thomas et al., 2013), bolide impacts (Birks et al., 2007), and rare but intense astrophysical ionizing radiation events such as gamma-ray bursts (Thomas et al. 2015) and supernovae (Gehrels et al., 2003).

Atmospheric $O_2$ has varied over geologic time; Harfoot et al. (2007) find that $O_3$ column densities are relatively insensitive to changes in $O_2$ experienced during the Phanerozoic, with values varying by about 30 DU or less due to $O_2$ variation.

Volcanic activity introduces a number of different constituents into Earth's atmosphere (Robock, 2000). Some are particularly important for transmission of UVB. Sulfur dioxide ($SO_2$) as a gas absorbs UVB, and is oxidized to sulfate aerosols ($H_2SO_4$) that further interact with UVB (McKeen and Liu, 1984; Dlugokencky et al., 1996). Ozone depletion is associated with volcanic activity; column density reductions of 5-20% have been observed and modeled following modern eruptions (Rosenfield et al., 1997; Robock, 2000). Continental flood basalt eruptions may have caused up to 70% depletion of $O_3$ (Black et al. 2014).

Many volcanic eruptions explosively eject constituents into the stratosphere, with plume heights reaching 30 km or higher (Mattis et al., 2010; Bruhl et al., 2015). $SO_2$ emitted is oxidized to form sulfate aerosols with a conversion rate of 30-40 days (McKeen and Liu, 1984; Mattis et al., 2010; Bruhl et al., 2015). Mixing between the stratosphere and troposphere is relatively slow, so constituents lifted above the tropopause are relatively long-lived (months to years), unless they are strongly removed by *in situ* processes such as photolysis. While ash is an important component of volcanic emissions, its residence time in the atmosphere is short, on the order of weeks (Mattis et al., 2010); we will not consider radiative effects of ash here.

Material introduced to the stratosphere at low latitudes can be lofted upward and transported poleward by Brewer-Dobson circulation, mixing throughout a hemisphere, or for equatorial events, globally. Atmospheric effects of material injected to the stratosphere by volcanoes are known to last for months to years (Robock, 2000; Mattis et al., 2010; Bruhl et al., 2015). Radiative heating due to the presence of sulfate aerosols can cause enhanced tropical upwelling, increasing transport to the stratosphere (Bruhl et al., 2015). Stratospheric temperatures can be increased, which reduces downward transport, leading to longer lived aerosol extinction (Rosenfield et al., 1997).

Recent volcanic activity has been dominated by occasional, mostly minor eruptions that affect the atmosphere for a few months to a year (Mattis et al., 2010; Bruhl et al., 2015). While large, explosive eruptions may have the greatest effects (for a single event), even small events may have atmospheric effects similar to much larger eruptions (Rampino and Self, 1984), especially if they occur in the tropics where upwelling can lift constituents into the stratosphere (Bruhl et al., 2015; Bourassa et al., 2012). During some periods in geologic history volcanic activity has been more intense and sustained. Continental flood basalt eruptions (also known as large igneous province eruptions) occur throughout geologic history (Wignall, 2005), with durations of a million years or more; prominent examples are the periods of volcanism that formed the Siberian (Renne and Basu, 1991) and Deccan Traps (Self et al., 2006). Such events can produce massive emission of $SO_2$ (and other constituents), for long periods of time, leading to a long-lived presence of $SO_2$ gas and sulfate aerosols (Self et al., 2006; Black et al. 2014; Schmidt et al. 2016). Modern day observations confirm that flood lava eruptions can produce large emissions of $SO_2$ (Schmidt et al., 2015). Rampino and Self (1984) argue that flood basalt eruptions may have a greater effect on Earth's atmosphere than any other volcanic event.

Major flood basalt eruptions are also likely to lead to significant $O_3$ depletion through the release of halogen species (HCl, $CH_3Cl$, $CH_3Br$), which deplete ozone in catalytic cycles (Wignall, 2005; Beerling et al., 2007; Black et al. 2014). Depletion of $O_3$ has been proposed as one aspect of the cause of the great end-Permian mass extinction (Beerling et al., 2007; Meyer and Kump, 2008).

Flood basalt eruptions may be particularly significant for a paleoaltitude proxy, since these periods of eruption are often associated with uplift (Hales et al., 2005; Rainbird, 1993), and Lomax et al. (2012) suggest uplift could be measured by their UVB/paleoaltitude proxy.

Clouds can also have a significant impact on UVB irradiance. This effect, however, is complicated and depends on the characteristics of the cloud cover (Whitehead et al., 2000). Stratus clouds tend to decrease UVB transmission, and can reduce irradiance at the ground by up to 75%. Cirrus clouds also reduce transmission. However, in partly cloudy conditions, cumulus clouds can actually increase surface irradiance by up to 30% for short periods due to the "cloud edge effect" caused by scattering. Over geologic history different cloud conditions may be associated with different climate regimes (Willis et al., 2009), so the influence of clouds is relevant to modeling variation in UVB. However, given the complexities and limitation of our 1D model, we do not attempt to model these effects.

## 3. Methods

### 3.1 Radiative transfer model

We use version 5.0 of the publicly available Tropospheric Ultraviolet and Visible (TUV) 1D radiative transfer model (http://cprm.acd.ucar.edu/Models/TUV/ ; Madronich and Flocke, 1997). The model has been extensively tested, and used in previous studies of UV changes due to volcanic activity (Dlugokencky et al., 1996) and atmospheric perturbations following astrophysical ionizing radiation events (Thomas et al., 2015).

TUV allows the user to modify a variety of parameter values, including date, time, location (latitude and longitude), altitude, and column density and vertical profile of $O_3$ and $SO_2$. Several aerosol optical parameters can be modified: total aerosol optical depth (AOD), single scattering albedo (SSA), Angstrom coefficient ($\alpha$), a wavelength-independent scattering asymmetry factor ($g$), and vertical profiles. The model requires input of AOD at 550 nm. The optical depth at a given wavelength is given by $\tau_1 = \tau_2(\lambda_2/\lambda_1)^{\alpha}$ where $\tau_i$ is the optical depth at wavelength $\lambda_i$ and $\alpha$ is the Angstrom coefficient. The default aerosol profile in TUV is for modern continental conditions, with AOD = 0.235, SSA = 0.99, $\alpha$ = 1.0, and $g$ = 0.61.

The model outputs include total UVB irradiance and irradiance weighted by various biological weighting functions. Here the output of interest is UVB and relative biological damage values using the weighting functions for erythema (McKinlay and Diffey 1987) and plant damage (Caldwell 1971). Note that the TUV model includes only relative values for these BWFs, rather than absolute biologically-weighted irradiance.

All of our modeling is completed at local noon. We fix the date and location, but vary solar zenith angle (SZA) for some model runs. Below we describe our choices of other parameter values.

### 3.2 Variation of ozone column density

The current column density of $O_3$ in the atmosphere is about 350 Dobson Units (1 DU = $2.69 \times 10^{20}$ molecules $m^{-2}$), but this value varies with latitude and season, ranging from about 200 DU (excluding values associated with the anthropogenic ozone "hole" over Antarctica), up to about 500 DU (Burrows et al. 1999). Over geologic time, $O_3$ concentration has ranged from about 0.5 to 1.5 times the current value (Harfoot et al. 2007).

Observations and modeling of recent volcanic activity shows $O_3$ column density reductions of 5-20% (Rosenfield et al., 1997; Robock, 2000). Much greater depletions have been proposed for flood basalt eruptions associated with formation of the Siberian Traps and the end-Permian mass extinction (Beerling et al., 2007; Black et al., 2014). Using a 2D atmospheric chemistry-transport model, Beerling et al. (2007) computed $O_3$ depletion between 10% and 60%, lasting throughout the several hundred thousand year eruption period that created the Siberian Traps. Using a 3D global chemistry-climate model, Black et al. (2014) find globally averaged $O_3$ reduction between 5% and 70% depending on details of assumed emissions, with locally higher reductions in some cases. For ongoing emissions depletion can be sustained in a steady-state. These findings may also be applicable to formation of the Deccan Traps and other flood basalt eruption periods, but could differ due to differences in particular geologic characteristics.

Astrophysical sources of high-energy ionizing radiation (including both photons and charged particles) can lead to significant and long-lived $O_3$ depletion (Gehrels et al., 2003; Thomas et al., 2013; 2015). Such events are likely on few hundred million year time scales and may be associated with mass extinctions (Melott and Thomas, 2009).

We consider changes in UVB at Earth's surface for $O_3$ column densities 100 - 525 DU. This spans the likely range of variability through the Phanerozoic (0.5 to 1.5 times the current value);

reductions to 100 DU or so may occur for some extreme events such as the end-Permian. For analysis of $O_3$ changes, we use the default altitude profile in TUV.

3.3 Variation of $SO_2$ and sulfate aerosols

Volcanic activity is a driver of atmospheric variability on long and short time scales. In the context of a paleoaltimetry proxy based on changes in UVB (or biologically-weighted) irradiance, we would like to consider a set of parameters that might characterize the state of atmospheric constituents affected by volcanic eruptions. Most important here are decreases in $O_3$ column density, increases in $SO_2$ column density, and production of sulfate aerosols from the emitted $SO_2$.

We consider two categories (Jones et al. 2016): 1) “stratospheric” eruptions (e.g. Pinatubo), which are short-lived but eject material into the stratosphere; and 2) longer lived but less energetic “flood” eruptions (e.g. Laki, Siberian Traps) or degassing events that emit large amounts of gases but do not explosively eject material above the troposphere. Geologically important events such as the volcanism that created the Siberian and Deccan Traps were primarily flood events, but may also have included stratospheric episodes (Schmidt et al. 2016). Therefore, we will also consider a combination of stratospheric and flood conditions to simulate this possibility.

For eruptions that reach the stratosphere the majority of $SO_2$ will be converted to sulfate aerosols over 30-40 days (McKeen and Liu, 1984; Mattis et al., 2010; Bruhl et al., 2015); the aerosols have a lifetime of 1-2 years in the stratosphere and can be globally distributed (Stevenson et al.

2003). For eruptions that do not penetrate the stratosphere, some $SO_2$ will be converted to aerosols, but as much as 70% may be deposited or removed by other mechanisms before forming aerosols, in which case the lifetime is of order days (Stevenson et al. 2003). However, large, long-lived episodes of volcanism may lead to a steady-state condition where $SO_2$ is "resupplied" as it is converted to aerosols or otherwise removed from the atmosphere (Stevenson et al. 2003).

Hoff (1992) gives $SO_2$ column densities between about 1 and 11 DU for the 1991 Pinatubo eruption. Loyola et al. (2008) examined $SO_2$ column densities for several South and Central American eruptions between 2000 and 2007, finding values between 1 and 3 DU; the associated eruptions were all much smaller than Pinatubo. Schmidt et al. (2015) report $SO_2$ column densities of 2 to 10 DU, with localized maxima up to 45 DU, associated with the 2014–2015 flood lava eruption at Bárðarbunga (Iceland). Overall, we have an extreme range from 1 to 45 DU, with most values lying between 1 and 10 DU.

Since the vertical distribution of atmospheric constituents can significantly affect absorption and scattering, we have adapted the default TUV profiles for $SO_2$ and aerosols to ones more appropriate for volcanic changes. We use the same profile for both $SO_2$ and sulfate aerosols; the two constituents should roughly track each other spatially since the sulfate aerosols are produced from the sulfur dioxide injected by the volcano. We use different altitude profiles for our two different volcanic cases. Our stratospheric profile is adapted from McKeen et al. (1984; see also Aquila et al., 2012), and peaks in the lower stratosphere, around 25 km. Our flood profile is constructed using observations in Ge et al. (2016) for "persistent degassing" activity, and from Stevenson et al. (2003) and Oman et al. (2006) for the 1783-1784 Laki flood eruption. In Figure 1A we show $SO_2$ profiles assumed for the two eruption types. In Figure 1B we show aerosol optical depth profiles; the "background" is that built into the TUV model, corresponding to modern "continental" conditions, and our two volcanic cases are superimposed on that background. The horizontal axes in Figure 1 have arbitrary units; the profiles are scaled to a total column density or optical depth.

According to DeLuisi (1997), the asymmetry factor (*g*) in the Pinatubo eruption was between 0.74 and 0.80. The default value in TUV is 0.61. In all volcanic cases we take $g = 0.74$.

According to Mattis et al. (2010) in their study of volcanic aerosols over Europe in 2008-2009 (following eruptions in Alaska) the Angstrom coefficient values ranged between 1.0 and 2.0; Holben et al. (2001) give a similar range at Mauna Loa, Hawaii between 1993 and 1999.

Takemura and Nakajima (2002) give single scattering albedo of 0.96 for sulfate aerosols in general (not specifically for volcanic conditions). Rosenfield et al. (1997) used a value of 0.989 for their modeling of the stratospheric effects of the Pinatubo eruption. DeLuisi (1997) gives a range from 0.94 to 0.99 for Pinatubo.

Finally, an important parameter is the aerosol optical depth (AOD). Bruhl et al. (2015) give a range of 0.028 to 0.52 at 530 nm, considering a large number of eruptions, with the highest value for Pinatubo in 1991. Mattis et al. (2010) give a range of 0.01 to 0.05 at 535 nm for a number of smaller eruptions. Holben et al. (2001) give a range of 0.01 to 0.03 at 500 nm over Mauna Loa between 1993 and 1999. Oman et al. (2006) give a range of 0.05 to 0.55 for the 1783-1784 Laki flood eruption, with values between 0.4 and 0.5 for about 3 months; it is not clear at which wavelength their AOD values are defined; it appears to be 500 nm. Using an Angstrom coefficient value of $\alpha$ = 1.5 (see above and section 3.1) we get a range of 0.01 to 0.58 at 550 nm (the wavelength value used in TUV for AOD).

While cloud parameters are included in TUV, the model is 1D and therefore cannot capture 3D effects such as the cloud edge effect. We therefore restricted our modeling to clear-sky conditions.

An advantage of a modeling study is that we can vary one parameter at a time, as contrasted with field observations, where many variables are important and may not be known for any given set of measurements. Before investigating the altitude effect emphasized by Lomax et al. (2012), we performed a number of model runs designed to determine the sensitivity of UVB irradiance (and relative biological impact values) to variations in $O_3$ and $SO_2$ column density and profile, and aerosol parameters. Our choices for ranges in these parameters are explained in Section 3.3. Our "base" set of parameter values assumes a solar zenith angle of 0°, 350 DU $O_3$, 0 DU $SO_2$, and the default TUV aerosol parameters (AOD = 0.235, SSA = 0.99, $\alpha$ = 1.0, and $g$ = 0.61). Initially, we varied one set of parameters at a time in order to gauge how significant changes in that parameter are for UVB irradiance and relative biological impacts (specifically erythema and plant damage, see Section 1).

While a detailed analysis of temporal and spatial variation is beyond the scope of this work, we did consider cases intended to simulate variation in time during an eruption; we have defined parameter sets for "early" and "late," for stratospheric, flood and the combination of both simultaneously, with parameter values given in Table 1. The cases presented are all for SZA = 30°. In the volcanic cases we varied the $O_3$ and $SO_2$ column densities, altitude profiles of $SO_2$ and aerosol optical depth, total aerosol optical depth and Angstrom coefficient ($\alpha$), holding SSA constant at 0.99 and $g = 0.74$. For some comparisons, we defined a "clean" baseline set with $O_3$ column density of 350 DU, $SO_2$ column density of 0 DU, AOD = 0.001 (at 550 nm), SSA = 0.99, $\alpha = 1.0$, and $g = 0.61$.

The main assumption in defining "early" versus "late" is that $SO_2$ column densities are high initially with conversion to aerosols and depletion of $O_3$ delayed. A higher aerosol load is assumed to have higher AOD and larger $\alpha$. Our stratospheric cases have higher AOD; $SO_2$ in the troposphere (ie. the flood cases) may not produce as much sulfate aerosol due to relatively rapid removal by other processes (Stevenson et al. 2003), while removal in the stratosphere is slow, so most $SO_2$ will be converted to aerosols.

In the context of a paleoaltimetry proxy based on spore and pollen chemistry/morphology the long-term (greater than ~ a few months), steady-state conditions are more likely to be relevant than short-term changes. We therefore defined cases assuming a steady emission of $SO_2$ with steady-state conversion to sulfate aerosols, accompanied by $O_3$ depletion. We assume a smaller AOD and decrease in $O_3$ for the flood case, since effects are likely to be greater in the troposphere than the stratosphere. However, we note that significant $O_3$ depletion was still modeled by Black et al. (2014) even for a non-stratospheric parameter set.

## 4. Results

Throughout this section and Section 5 we use percent difference comparisons. These are percent differences between model output values given different parameter input values (e.g. two different $O_3$ column densities). This comparison is used, rather than absolute differences, in order to 1) examine relative changes, which are more demonstrative of the significance of a given parameter variation; and 2) enable direct comparison with other work (see Section 5).

### 4.1 Sensitivity studies

Figures 2 and 3 show percent differences between UVB, erythema and plant damage at different values of altitude, $O_3$ column density, $SO_2$ column density, aerosol parameter values, and altitude profile sets (note the difference in scale between Figures 2 and 3). Since solar zenith angle has a major effect on transmission of UVB we have performed this sensitivity study at two SZA values (0° and 30°). A higher solar zenith angle produces smaller irradiance values (Figure 5B) due to a longer column of atmosphere to travel through. However, the relative change in UVB (and relative biological damage) caused by other parameter changes (Figures 2 and 3) is very similar for these two SZA values.

Variations in $O_3$ have a major impact (Figure 2) on changes in UVB and biological effects. Figure 4 shows variation of UVB irradiance and relative biological impacts at sea level as a function of $O_3$ column density for SZA = 0°. A 25 DU decrease in $O_3$ column density can give around 10% reduction in UVB irradiance, with greater change in UVB as column densities decrease. At high column density values UVB is already strongly absorbed and so increases in $O_3$ have a diminishing effect.

4.2 Variation with altitude

We used TUV to compute UVB irradiance over a range of altitudes from sea level to 5 km (the Tibetan plateau has an average height of 4.5 km). We see in Figure 5 that there is indeed an increase in UVB irradiance with altitude, though this increase is not linear. Note that changes in $O_3$ have a strong impact on UVB irradiance received, at any altitude (over the range considered here). Figure 5B shows results for a single $O_3$ column density (350 DU) at three different SZA values (note the log scale in panel B). The SZA value has a significant impact on UVB irradiance, especially when the sun is very low in the sky, due to the long path length through the atmosphere.

The quantity of interest in Lomax et al. (2012) and similar work is the change in UVB (or biologically-weighted) irradiance with changes in altitude. Figures 6 and 7 show the percent difference per kilometer between model outputs at 1 km compared to 0 km, 2 km compared to 1 km, etc. (the "altitude effect"), for a number of different parameter variations. In Figure 6, for a wide range of SZA and $O_3$ column densities, we find that irradiance and relative biological damage increases by about 12-15% in the first kilometer (comparing sea level to 1 km altitude). As altitude is increased successively by 1 km, the percent difference per km becomes smaller, ranging from about 4% to about 6% between altitudes of 4 and 5 km. This result is not surprising since the density of the atmosphere decreases exponentially. While $O_3$ is a primary atmospheric absorber of UVB, scattering also plays an important role in determining irradiance and depends on air density as well as the constituents present (such as aerosols). Each additional km increase in altitude effectively gives a smaller change in overhead atmosphere leading to a decreasing change in UVB with altitude. Even a large change in SZA does not have a major impact on the altitude effect for UVB and biological damage with altitude (Figure 6, panels A, C, E). All our results show the same general trend, with small differences between UVB irradiance, erythema and plant damage, due to the wavelength-dependent biological response.

### 4.3 Effects of volcanic activity

As described in Section 2, volcanic activity leads to changes in a number of atmospheric constituents. The stratospheric and flood cases described in Section 3.3 use profiles for $SO_2$ and aerosols shown in Figure 1, and the combination case uses profiles that are simply the combination of the stratospheric and flood profiles. Results presented in Figures 2 and 3 give a sense of which parameters are most important and how variations in these parameters affect the UVB irradiance and relative biological impacts. The most important parameter is $O_3$ column density, followed by altitude change, AOD, and $SO_2$ column density. Other parameters have a small effect (about 5% or less).

Table 2 gives the resulting UVB irradiance and relative biological damage values for these cases, and Table 3 gives some comparisons between cases (in terms of percent differences). We find large differences in UVB irradiance and relative biological damage when comparing between cases and against a “clean” background atmosphere, indicating the importance of atmospheric changes caused by volcanic activity.

Figure 7 shows the percent difference per kilometer in altitude (as in Figure 6) for UVB and relative biological damage values for the cases presented in Tables 1-3. Compared to other parameters (Figure 6), we find a greater difference in altitude effect when comparing different volcanic cases (Figure 7). Here we find the increase per km in the first km varies between about 9 and 16%; this is a much larger range of variation than the 12-15% when changing only $O_3$ and SZA (Figure 6).

## 5. Discussion

In this section we examine two primary results: 1) how the "altitude effect" is affected by variations in parameters considered in this work; 2) how absolute values of UVB irradiance and relative biological damage are affected by changes in those same parameters. The altitude effect is important for the proposed altitude proxy, while absolute values are important for investigations of the UV environment at Earth's surface over geologic time.

### 5.1 Variation with altitude

Past work has found that UVB irradiance (and in some cases biologically-weighted irradiance) varies by 5-20% per 1 km elevation (Lomax et al. 2012 and references therein). This range represents variation due to differing conditions under which irradiance was measured. Blumthaler et al. (1997) measured erythema-weighted irradiance at 3576 m above sea-level (a.s.l.) and at 577 m a.s.l. and found variation of 18% ± 2% per 1000 m. Gonzalez et al. (2007) measured irradiance at 4300 and 5000 m a.s.l. in summer and winter and found a mean altitude effect of 13.5% per 1000 m. Sola (2008) found 7-11% per 1000 m. Dubrovsky (2000) measured irradiance at 278 and 827 m a.s.l. and found only 4-8% per 1000 m. Of the studies available, most cluster around 12-18% per 1000 m, with Dubrovsky (2000) being an outlier. Our results (Section 4.2) for the first 1 km increase from sea level are in fairly good agreement with these measurements when varying $O_3$ and SZA (Figure 6). On the other hand, our “Straospheric” and “Combination” volcanic cases (Figure 7) show larger change in UVB (and biologically-weighted irradiance), especially in the first 1-3 km. This is not surprising since the observational data were not collected under these volcanic conditions. These results do point out the importance of atmospheric changes associated with volcanic activity.

While much of the work discussed above and used in Lomax et al. (2012) refers to the “altitude effect” wherein change in irradiance is given per additional km of altitude, Lomax et al. (2012) in their Figure 1 also show percent increase in erythemal UVB flux as a function of altitude compared to sea level, for several geographic regions. In Figure 8 we compare our results using this same measure to values from Lomax et al. (2012) Figure 1. (See also our Figure 2.) For a wide range of parameters considered in this work, the percent increase in erythema compared to sea level is fairly consistent, though the Pacific and North America values given by Lomax et al. (2012) are smaller and larger, respectively, and our “combination” stratospheric plus flood volcanic case is larger than our other cases above about 2 km.

### 5.2 Variation of absolute values

Figures 2, 3, 4 and 5 allow us to evaluate how absolute values of UVB irradiance and relative erythema and plant damage are affected by various parameters. First, change in $O_3$ column density has a major impact, as expected. Variation in $O_3$ column density alone clearly introduces significant uncertainty into any attempt to tie UVB irradiance variation to altitude change, if comparing spore/pollen samples that grew under different $O_3$ column densities. A 25 DU decrease in $O_3$ column density can give around 10% reduction in UVB irradiance, mimicking a 1 km altitude change; such a change is similar in magnitude to that associated with varying $O_2$ levels through the Phanerozoic (see Section 2). Seasonal and latitudinal differences of 100-200 DU or more can give UVB irradiance changes of 25-35%.

A higher solar zenith angle produces smaller irradiance values (Figure 5B). Therefore, annual seasonal changes in sun angle and differences in latitude will cause potentially large changes in UVB and biological-weighted irradiance. On the other hand, differences in UVB and biological damage under variations of other parameters are similar at two different SZA values (Figures 2 and 3).

Figures 2 and 3 show that after $O_3$ column density, aerosol optical depth (AOD) and $SO_2$ column density are the next most important variables. Each can give a reduction in UVB irradiance or relative biological damage of 10-15%, again mimicking a 1 km altitude change, if comparing spore/pollen samples that grew under different AOD or $SO_2$ conditions.

Changes in AOD and $SO_2$ (as well as $O_3$) are important in the context of varying volcanic activity. For the cases detailed in Table 1 with results in Table 2, we find large differences (Table 3) in UVB irradiance and relative biological damage when comparing between cases and against a "clean" background atmosphere. These differences in some cases are far larger than changes caused by any of the individual variables discussed above. It is interesting to note that the differences are largest for plant damage, followed by erythema, with significantly smaller (though still large) differences in UVB. This points out the importance of the wavelength dependence of biological impacts.

Schmidt et al. (2016) simulate AOD values between 0.5 and 2.0 (at 550 nm) for a Deccan-scale flood basalt eruption. While we did not include these very high values in our detailed modeling, setting AOD = 2.0 (holding all other variables at their default values) decreases UVB irradiance by about 40%, with a similar decrease in erythema and plant damage. This again points out the importance of including all the relevant parameters in modeling UVB exposure during volcanic episodes.

## 6. Conclusions

This study provides important data for researchers interested in assessing changes in the UV environment at Earth's surface through geologic time, especially during periods of major volcanic activity, as well as attempts to measure paleoaltitude using UVB proxies. We have identified a number of factors that must be considered, and our results have important implications for comparing samples across space and time. If samples from a volcanically active region or time are compared to samples from a less active region or time then differences in UVB irradiance indicated by the spore/pollen proxy will not accurately indicate altitude difference. A similar conclusion follows for samples from regions or times that differ in $O_3$ column density or solar zenith angle.

On the other hand, an important result of this work is that the change in irradiance per kilometer (Figures 5, 6 and 7) and versus sea level (Figure 8) with increasing altitude is fairly consistent across a wide range of parameter differences, and are certainly within the variation of modern measurements taken in different regions and under different atmospheric conditions (Figure 8).

This indicates that if samples from a given region and time period are compared to each other, conclusions about relative altitude should be reliable, since variation with altitude for one set of parameters is not greatly different than for a different set of parameters.

Our results are also useful for studies seeking to understand the UV environment at Earth's surface over geologic time, especially during mass extinctions. For instance, under conditions likely associated with the formation of the Siberian Traps, and possibly the Deccan Traps and other flood basalt eruption periods, the UVB irradiance could span a wide range (Tables 2 and 3). While further analysis of particular geologic events is beyond the scope of this study, we do note that this result has implications for claims that heightened UVB irradiance contributed to the end-Permian mass extinction. Previous studies that considered the increase in UVB due only to depletion of $O_3$ (e.g. Black et al. 2014) fail to include the important effects of $SO_2$ and aerosols, and therefore may not accurately capture the ground-level UVB irradiance and subsequent biological implications. On the other hand, we do find (Table 3) that UVB irradiance (and relative biological impact) is increased in all cases where significant $O_3$ depletion is assumed.

Over geologic time, $O_2$ in Earth's atmosphere has varied, but Harfoot et al. (2007) find that $O_3$ column densities are relatively insensitive to $O_2$ changes during the Phanerozoic, with values varying by about 30 DU. According to our results this corresponds to a difference in UVB irradiance and relative biological damage of about 10%.

It should be noted that our study did not consider effects of clouds, which could be significantly different during volcanically active periods and can have major effects on UVB transmission; future studies could be done to better quantify this uncertainty, but care must be taken with the complicated microphysics of clouds and their 3D effects on radiative transfer.

While we have attempted to use the best available estimates of conditions associated with geologically important events such as the volcanism that formed the Siberian Traps, much uncertainty exists in all the parameters evaluated. The best modeled constituent change associated with this particular event appears to be variation in $O_3$ (e.g. Black et al. 2014), which in any case is the most significant variable considered here in changing UVB irradiance at the surface.

Any attempt to apply a proxy based on changes in UVB (inferred by any method) across location and time must take into account a wide range of highly variable atmospheric factors. The roughly 10-15% change in UVB due to a 1 km change in altitude can easily be mimicked or overwhelmed by realistic changes in atmospheric constituents, on both short and long time scales.

In order to apply the proxy, one must be confident that the samples under comparison were located at roughly the same latitude (so that the SZA does not differ significantly), under very similar $O_3$ columns (within about 25 DU difference), with similar atmospheric conditions (especially aerosol properties and $SO_2$ column). Holding these variables constant between the two samples may allow relative altitudes to be deduced within about 1 km as suggested by Lomax et al. (2012), though precision of less than ±1 km seems unrealistic.

We draw these general conclusions: 1) UVB irradiance and associated biological impacts are strongly affected by sun angle (ie. season, latitude), $O_3$ and $SO_2$ column density, and aerosol optical depth; 2) UVB-proxy properties will be significantly affected by differences in these quantities. Overall, accurate estimates of the surface-level UVB exposure at any time and location must take into account a number of atmospheric factors. This has important implications not only for paleoaltitude proxies, but also for studies of the UV environment over time and attempts to tie extinctions to UVB irradiance changes.

Acknowledgements

This work has been supported by the National Aeronautics and Space Administration under grant No. NNX14AK22G, through the Astrobiology: Exobiology and Evolutionary Biology Program. Computational time for this work was provided by the High Performance Computing Environment (HiPACE) at Washburn University; thanks to Steve Black for assistance with computing resources. Thanks to Robert J. Gutro, Ellen T. Gray, Aquila Valentina, Jean Paul Vernier, Anja Schmidt and Benjamin D. Santer for assistance with data on volcanic aerosols. Thanks to Adrian Melott and Bruce Lieberman for helpful comments. We thank the referees, whose comments helped to significantly improve the paper.

Author Disclosure Statement: No competing financial interests exist.

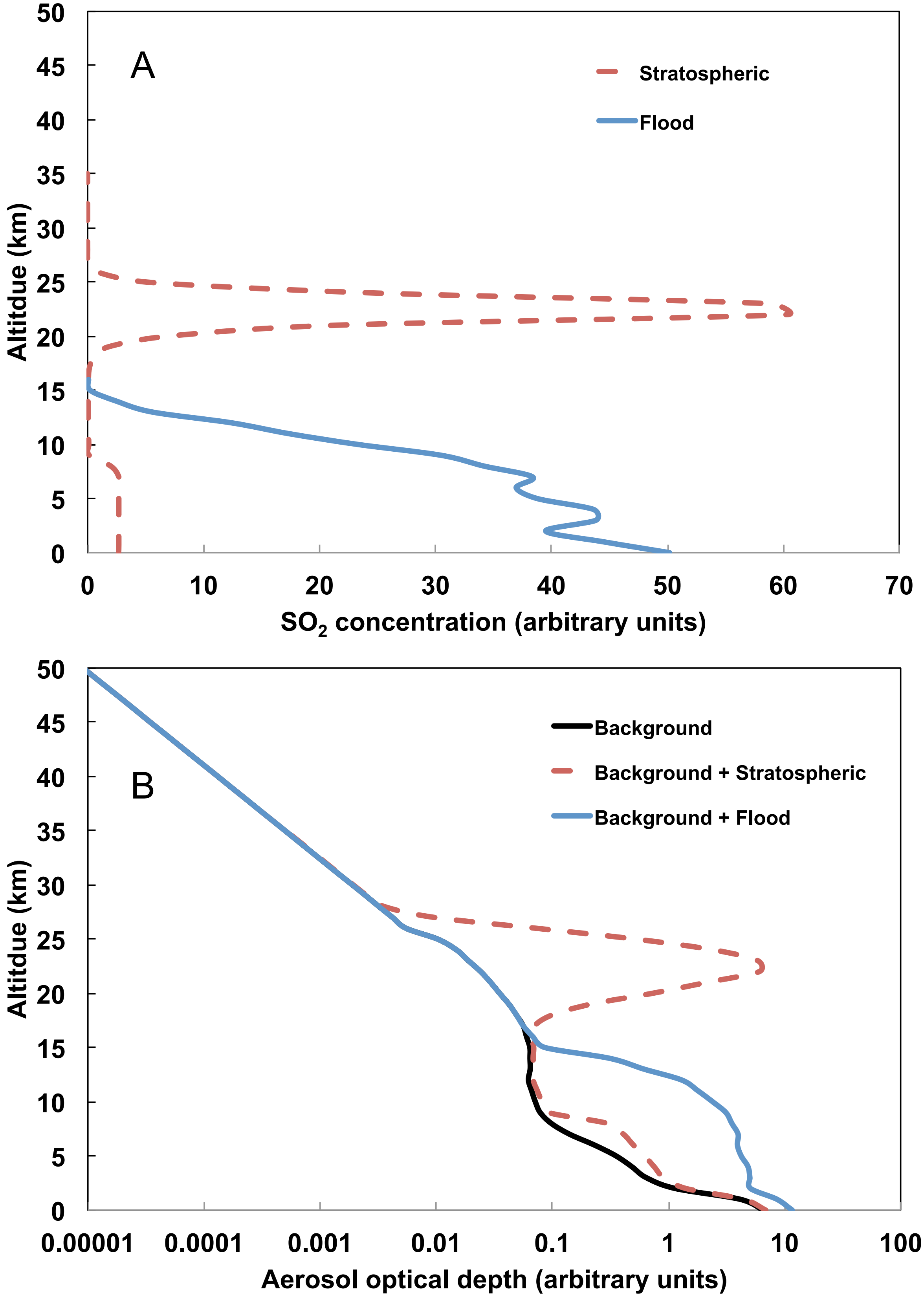
A
Stratospheric
Flood
Altitdue (km)
SO2 concentration (arbitrary units)
0
10
20
30
40
50
60
70
0
5
10
15
20
25
30
35
40
45
50
B
Background
Background + Stratospheric
Background + Flood
Altitdue (km)
Aerosol optical depth (arbitrary units)
0.00001
0.0001
0.001
0.01
0.1
1
10
100

Figure 1

Vertical profiles of $SO_2$ concentration (panel A) for our two volcanic cases and aerosol optical depth (panel B) for the two volcanic cases and the default TUV background. Note horizontal axes have arbitrary units since values are scaled in TUV to a given column density or total optical depth. Differing altitude distribution of both $SO_2$ and AOD affects transmission of UVB to the surface. In the stratospheric case material is ejected above the tropopause, while in the flood case the distribution is limited to the lower atmosphere. The stratospheric profile is adapted from McKeen et al. (1984) and Aquila et al. (2012). The flood profile is adapted from Ge et al. (2016), Stevenson et al. (2003) and Oman et al. (2006).

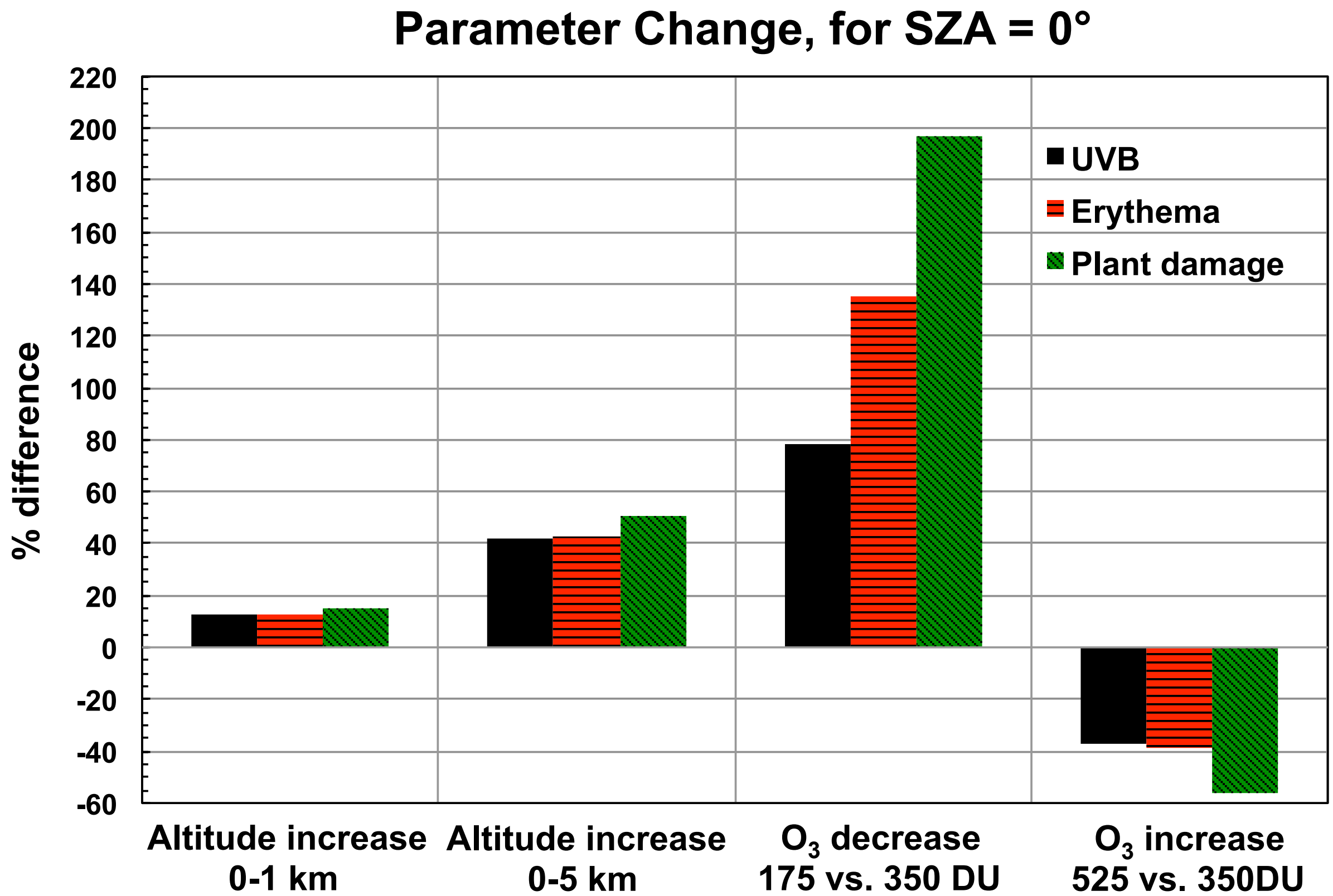
Parameter Change, for SZA = 0°
220
200
180
160
140
120
100
80
60
40
20
0
-20
-40
-60
% difference
UVB
Erythema
Plant damage
Altitude increase 0-1 km
Altitude increase 0-5 km
O3 decrease 175 vs. 350 DU
O3 increase 525 vs. 350DU

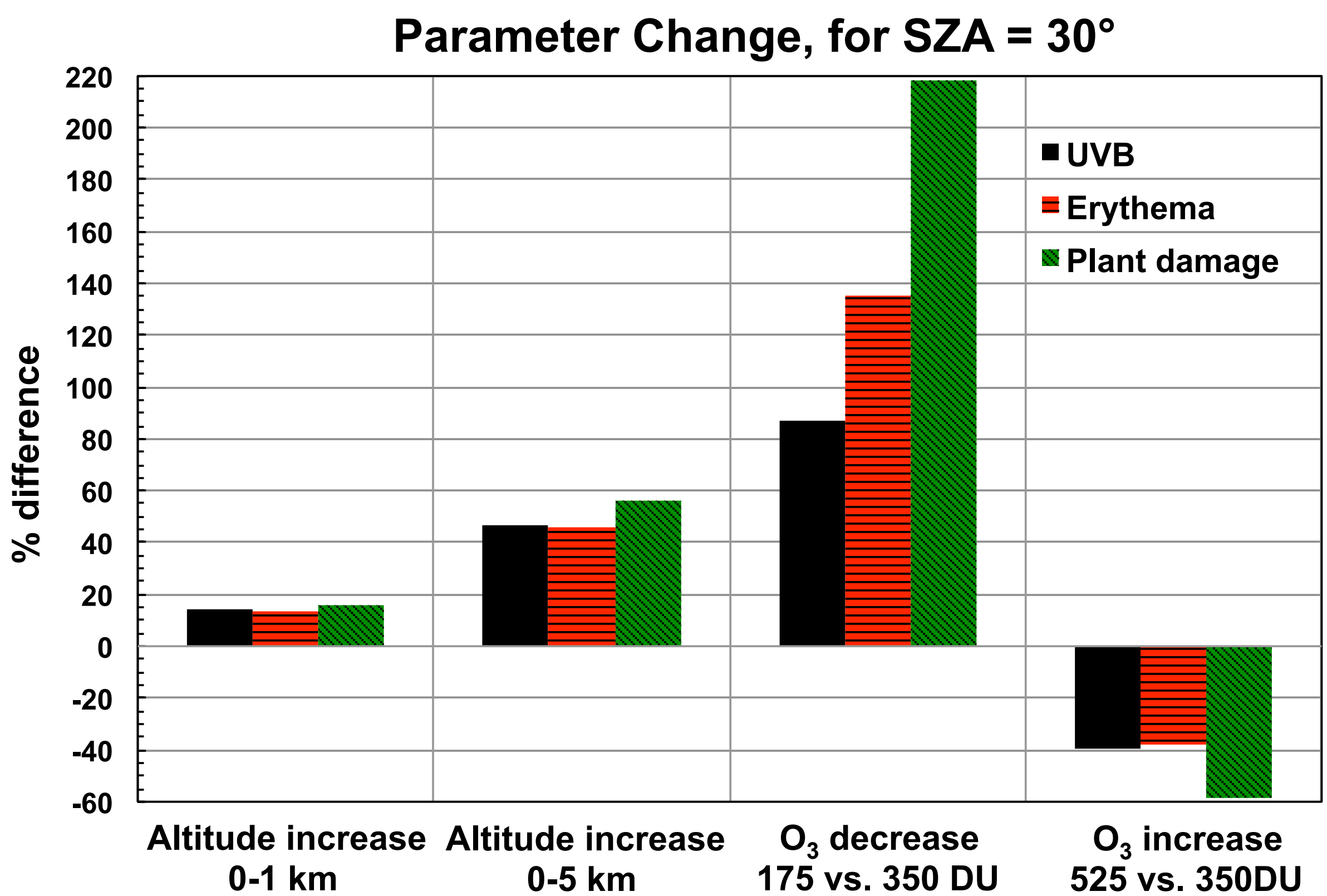
Parameter Change, for SZA = 30°
220
200
180
160
140
120
100
80
60
40
20
0
-20
-40
-60
% difference
UVB
Erythema
Plant damage
Altitude increase 0-1 km
Altitude increase 0-5 km
O3 decrease 175 vs. 350 DU
O3 increase 525 vs. 350DU

Figure 2

Percent differences, at SZA 0° and 30°, of UVB irradiance, relative erythema and relative plant damage, comparing values at altitude differences 0 to 1 km and 0 to 5 km; $O_3$ decrease from 350 DU to 175 DU; $O_3$ increase from 350 to 525 DU. These ranges of $O_3$ column density span the extremes likely to have existed throughout the Phanerozoic. Changes in $O_3$ are dominant over other parameter variations here and in Figure 3. In all cases here plant damage shows the largest change.

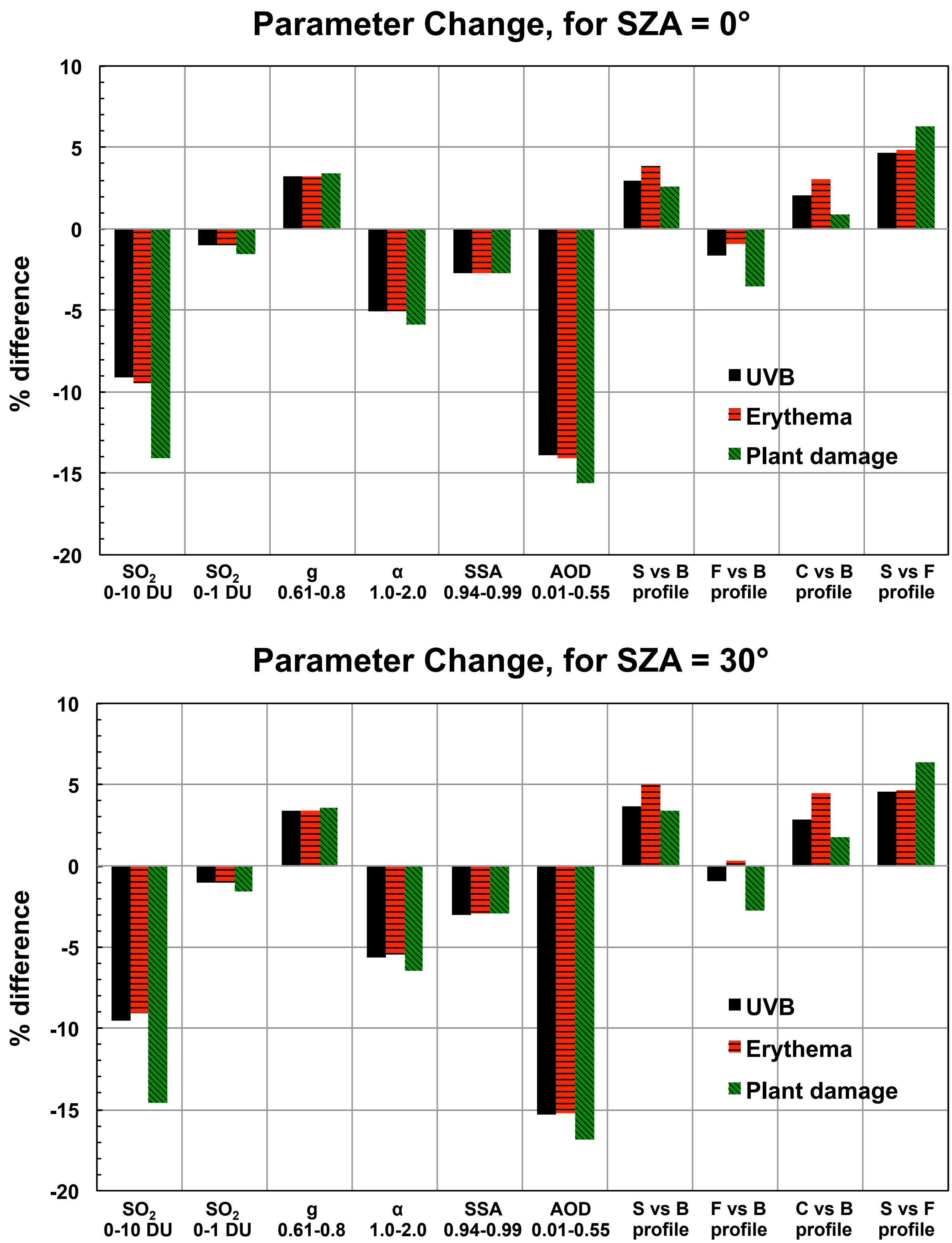
Parameter Change, for SZA = 0°
% difference
10
5
0
-5
-10
-15
-20
UVB
Erythema
Plant damage
SO2 0-10 DU
SO2 0-1 DU
g 0.61-0.8
α 1.0-2.0
SSA 0.94-0.99
AOD 0.01-0.55
S vs B profile
F vs B profile
C vs B profile
S vs F profile
Parameter Change, for SZA = 30°
% difference
10
5
0
-5
-10
-15
-20
UVB
Erythema
Plant damage
SO2 0-10 DU
SO2 0-1 DU
g 0.61-0.8
α 1.0-2.0
SSA 0.94-0.99
AOD 0.01-0.55
S vs B profile
F vs B profile
C vs B profile
S vs F profile

Figure 3

As in Figure 2, percent differences at SZA 0° and 30° of UVB irradiance, relative erythema and relative plant damage. Here, comparing values for increases of $SO_2$ from 0 to 10 DU and 0 to 1 DU; aerosol asymmetry factor (*g*) 0.61 to 0.8; Angstrom coefficient ($\alpha$) 1.0 to 2.0; aerosol single scattering albedo (SSA) 0.94 to 0.99; aerosol optical depth (AOD) 0.01 to 0.55; and for the $SO_2$ and aerosol profiles shown in Figure 1, where "S" is stratospheric profile, "F" is flood profile, and "C" is combined stratospheric and flood profile; these profiles are compared to the "B" base (default) TUV profile, and the stratospheric and flood profiles are compared to each other. Here, the increase in AOD gives the largest change in model results, followed by a 0-10 DU increase in $SO_2$. Comparison of plant damage, erythema and un-weighted UVB irradiance is more complicated here than in Figure 2, in particular when comparing different altitude profiles of $SO_2$ and AOD. This is due to the more complicated wavelength dependence of scattering combined with absorption, versus the more simple absorption-only change with changes in $O_3$.

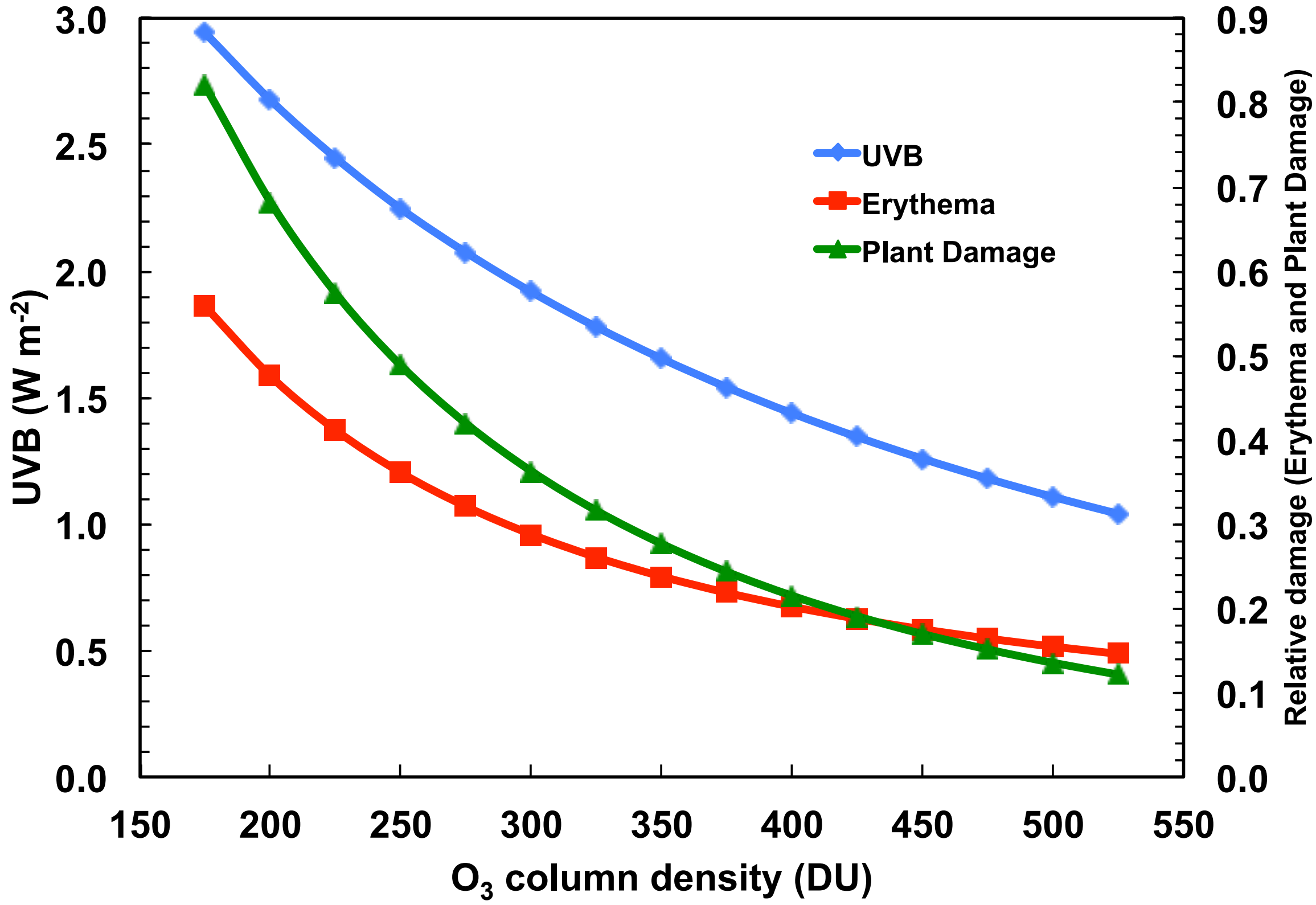

Figure 4

UVB irradiance (left vertical axis) and relative erythema and plant damage (right vertical axis) as a function of $O_3$ column density, at SZA = 0°. This figure presents a more detailed view of how changes in $O_3$ column density affect changes in model results. The rate of change is greater at smaller $O_3$ values, meaning an increase/decrease around smaller values has a greater impact on irradiance and biological damage. This difference is greater for plant damage than erythema.

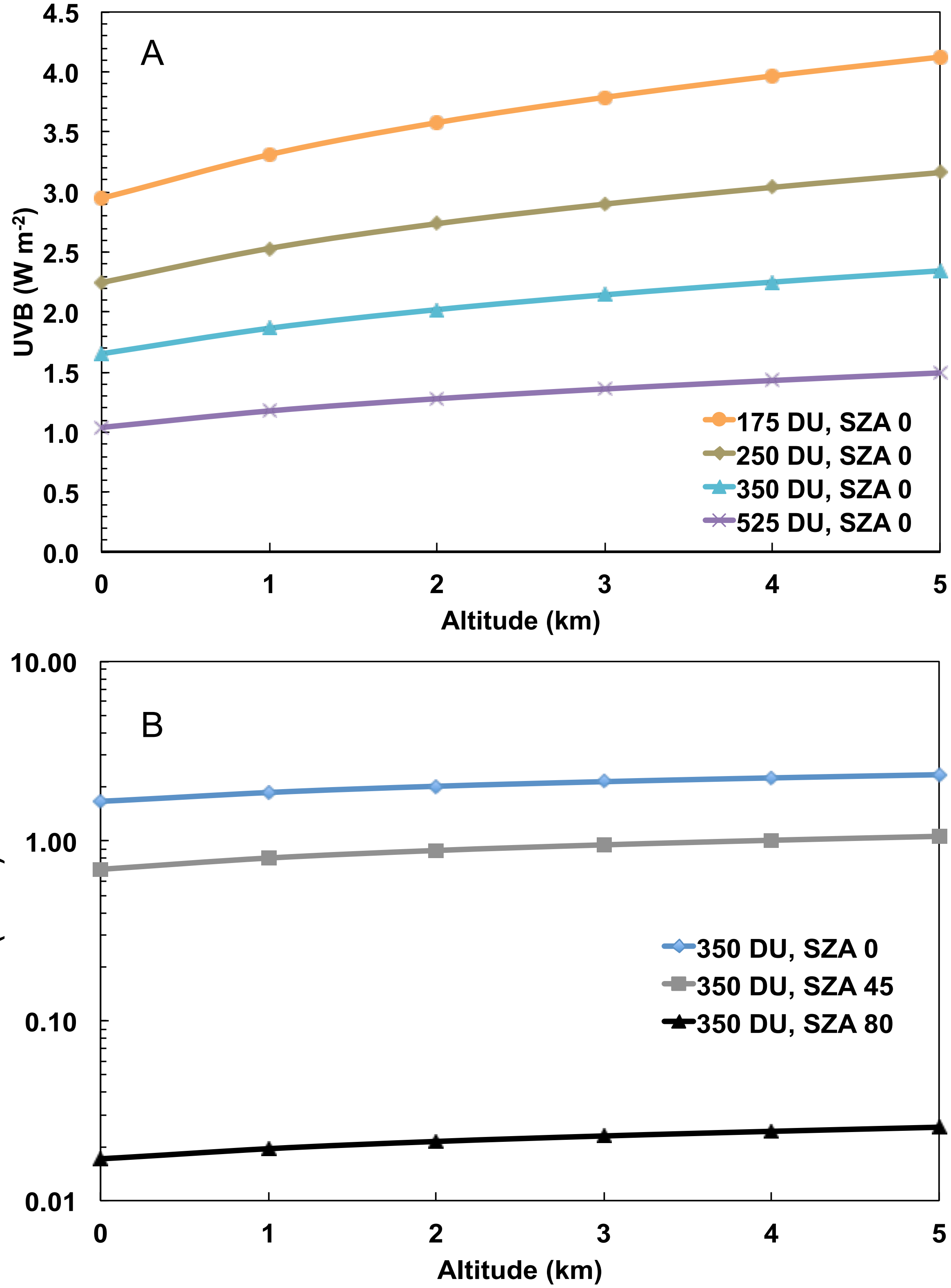
A
UVB (W m-2)
4.5
4.0
3.5
3.0
2.5
2.0
1.5
1.0
0.5
0.0
0
1
2
3
4
5
Altitude (km)
175 DU, SZA 0
250 DU, SZA 0
350 DU, SZA 0
525 DU, SZA 0
B
UVB (W m-2)
10.00
1.00
0.10
0.01
350 DU, SZA 0
350 DU, SZA 45
350 DU, SZA 80

Figure 5

Variation of UVB irradiance with altitude for several $O_3$ column density values (175, 250, 350 and 525 DU) all at SZA = 0° (panel A) and at three different solar zenith angle values (0°, 45°, and 80°) for $O_3$ column density 350 DU (panel B). Note the log vertical scale in panel B due to the large difference in irradiance at SZA = 80° vs. 0°. Change in UVB irradiance is fairly consistent over the altitude range, but greatest in the lowest 1 km, due to the exponential decrease of air density with altitude.

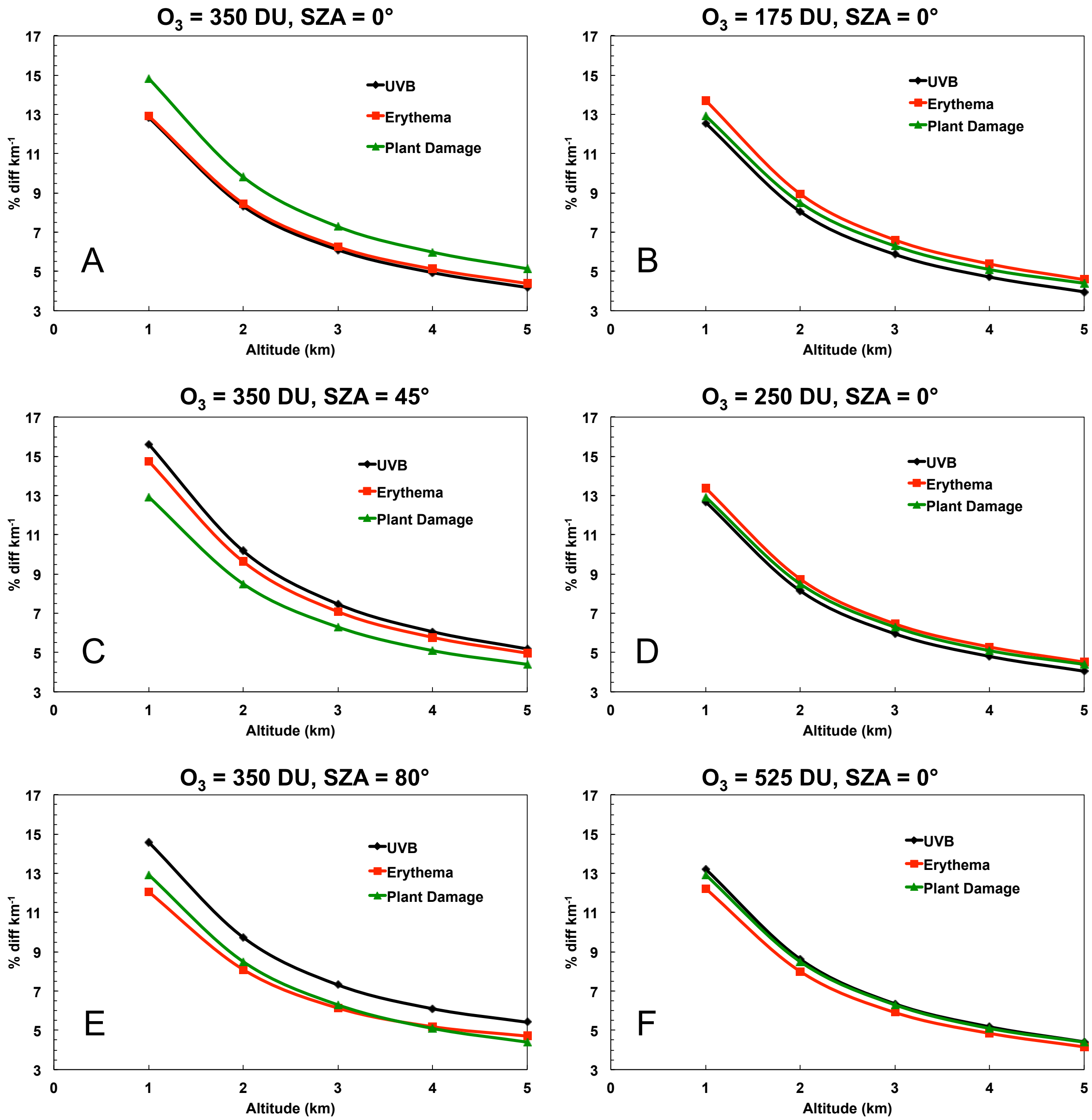

Figure 6

Percent difference per km as a function of altitude, comparing values of UVB irradiance and relative erythema and plant damage at 1 km to 0 km, 2 km to 1 km, etc., for $O_3$ column density 350 DU at SZA = 0°, 45°, and 80° (panels A, C and E) and $O_3$ column densities of 175 DU (panel B), 250 DU (panel D) and 525 DU (panel F) at SZA = 0°. As in Figure 5, the difference decreases with altitude, again due to the exponential decrease in air density. All cases show a similar trend with altitude, and in most cases the three model outputs examined track each other rather closely.

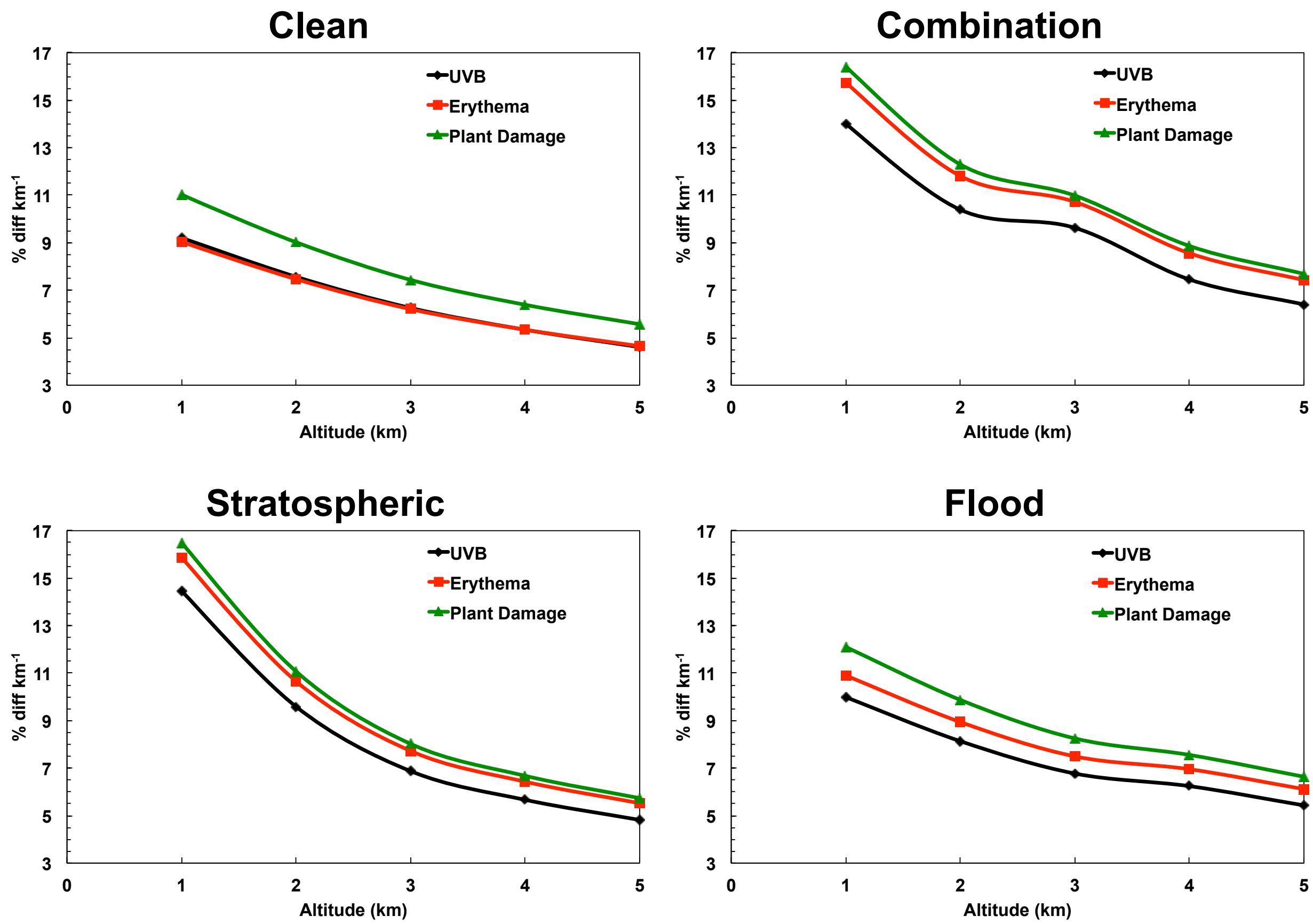

Figure 7

As in Figure 6, percent difference per km as a function of altitude, comparing values of UVB irradiance and relative erythema and plant damage at 1 km to 0 km, 2 km to 1 km, etc., for cases detailed in Table 1 (a “clean” atmosphere, “stratospheric” volcanic case, “flood” volcanic case, and “combination” of flood and stratospheric volcanic cases). The results here are much more variable between cases, when compared with Figure 6. As noted in Figure 3, variation in altitude distribution of constituents introduces more complicated changes in surface-level irradiance, leading to this greater variation. However, UVB irradiance, erythema and plant damage still track each other fairly closely.

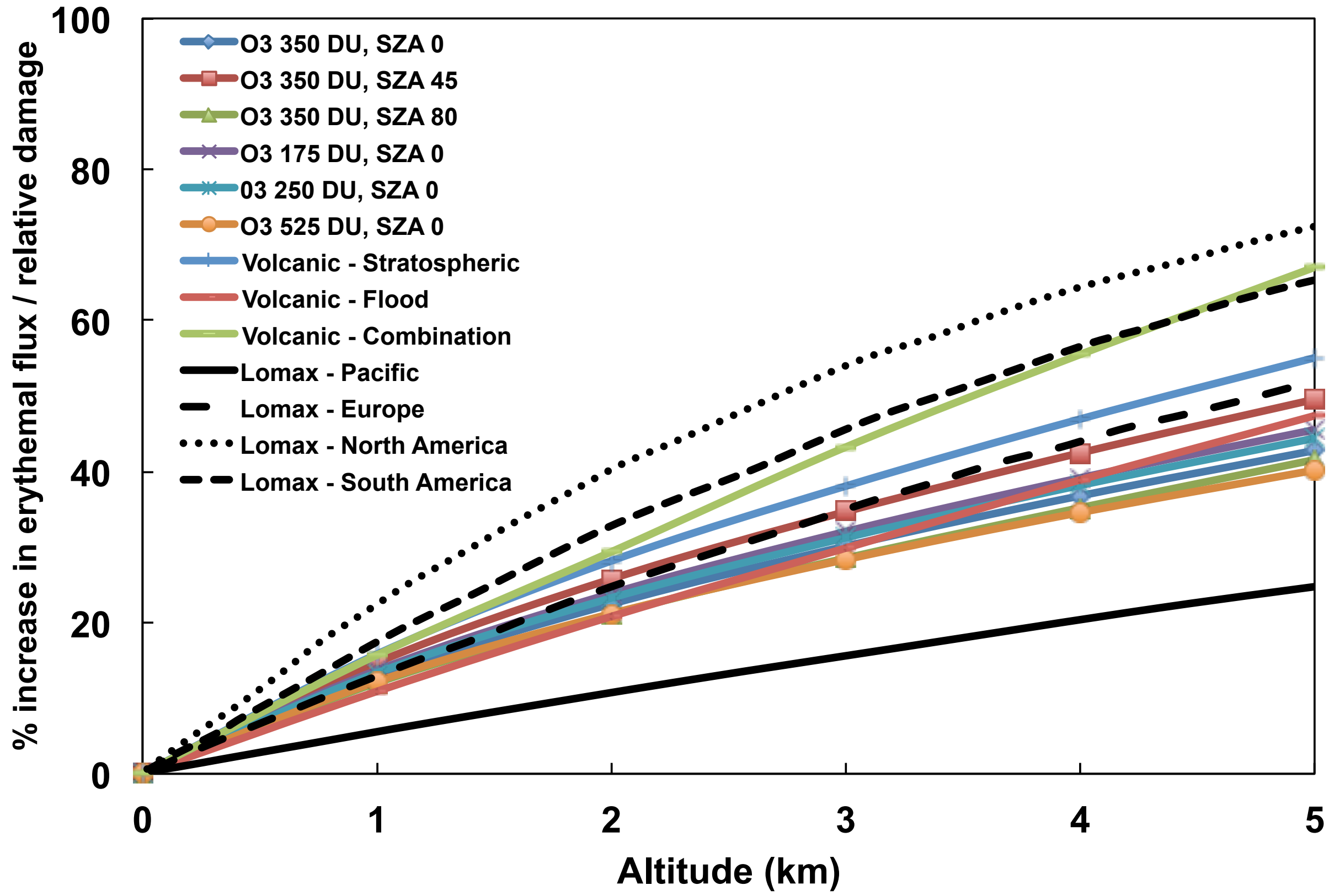

Figure 8

Percent increase in erythemal flux (from Figure 1 of Lomax et al., 2012) and relative erythema damage values (other cases, this work) as a function of altitude, comparing values at each altitude to those at 0 km (sea level). Cases include those considered in this work, for $O_3$ column densities 175, 250 and 525 DU at SZA = 0°; $O_3$ column density 350 DU at SZA = 0°, 45°, and 80°; and stratospheric, flood and combination volcanic cases defined in Table 1, as well as data from Figure 1 of Lomax et al. (2012) for four different geographical regions (Pacific, Europe, North America, South America). A large number of cases are included here in order to allow for comparison between our different results as well as those from Lomax et al. (2012). In most of our cases the results track fairly closely together, and all our cases fit within the range of measurements from Lomax et al. (2012). As in Figure 7, our “combination” case is an outlier, at least at higher altitudes, but is still within the Lomax range.

Table 1

Parameter values for a "clean" atmosphere and several volcanic cases used to generate surface-level UVB irradiance and relative biological impact values. All cases are for SZA = 30°.

| Case | $O_3$ (DU) | $SO_2$ (DU) | AOD | $\alpha$ |
|---|---|---|---|---|
| Clean base | 350 | 0 | 0.001 | 1.0 |
| Stratospheric, early | 350 | 10 | 0.010 | 1.0 |
| Stratospheric, late | 100 | 1 | 0.500 | 2.0 |
| Flood, early | 350 | 10 | 0.010 | 1.0 |
| Flood, late | 200 | 1 | 0.050 | 1.0 |
| Combination, early | 350 | 10 | 0.010 | 1.0 |
| Combination, late | 100 | 1 | 0.500 | 2.0 |
| Stratospheric, steady | 100 | 5 | 0.500 | 2.0 |
| Flood, steady | 200 | 5 | 0.050 | 2.0 |
| Combination, steady | 100 | 5 | 0.500 | 2.0 |

Table 2

UVB irradiance and relative biological damage values at sea level computed using parameter values detailed in Table 1. All cases are for SZA = 30°.

| Case | UVB (W $m^{-2}$) | Erythema | Plant Damage |
|---|---|---|---|
| Clean base | 1.250 | 0.176 | 0.187 |
| Stratospheric, early | 1.090 | 0.152 | 0.157 |
| Stratospheric, late | 2.600 | 0.582 | 0.890 |
| Flood, early | 1.080 | 0.151 | 0.155 |
| Flood, late | 2.050 | 0.341 | 0.471 |
| Combination, early | 1.070 | 0.149 | 0.154 |
| Combination, late | 2.630 | 0.601 | 0.924 |
| Stratospheric, steady | 2.420 | 0.531 | 0.809 |
| Flood, steady | 1.910 | 0.313 | 0.428 |
| Combination, steady | 2.430 | 0.540 | 0.825 |

Table 3

Percent differences in UVB irradiance and relative biological damage for volcanic cases detailed in Table 1, using results in Table 2. Values labeled "late v. early" are comparison between "early" and "late" case results in Table 2. Values labeled "v. clean" are comparison between volcanic case and clean atmosphere case.

| | UVB | | Erythema | | Plant Damage | |
|---|---|---|---|---|---|---|
| Case | late v. early | v. clean | late v. early | v. clean | late v. early | v. clean |
| Stratospheric, early | -- | -13 | -- | -13 | -- | -16 |
| Stratospheric, late | 139 | 108 | 283 | 232 | 467 | 377 |
| Flood, early | -- | -14 | -- | -14 | -- | -17 |
| Flood, late | 90 | 64 | 126 | 94 | 204 | 152 |
| Combination, early | -- | -14 | -- | -15 | -- | -18 |
| Combination, late | 146 | 110 | 303 | 242 | 500 | 395 |
| | | | | | | |
| Stratospheric, steady | -- | 94 | -- | 202 | -- | 333 |
| Flood, steady | -- | 53 | -- | 78 | -- | 129 |
| Combination, steady | -- | 94 | -- | 208 | -- | 342 |